\documentclass[a4paper,11pt]{article}
\usepackage{pos}

\usepackage{physics}

\usepackage{setspace}
\usepackage[T1]{fontenc}
\usepackage{soul}
\usepackage{graphicx}
\usepackage{lipsum,multicol}
\usepackage{cancel}
\usepackage{fancyhdr}
\usepackage{pgf,tikz}
\usetikzlibrary{arrows}
\usepackage{url}
\usepackage{chemfig}
\usepackage{braket}

\usepackage{gensymb}
\usepackage{mathrsfs}
\usepackage{mathtools}
\usepackage{xspace}
\usepackage{multirow}
\usepackage{afterpage}
\usepackage{pdflscape}
\usepackage{arydshln}
\usepackage[normalem]{ulem}
\usepackage{lineno}
\usepackage[bottom]{footmisc}
\usepackage{subfigure}
\usepackage[toc,page]{appendix}
\usepackage{sidecap}
\usepackage{wrapfig}
\usepackage{stmaryrd}
\usepackage{float}
{%
\stepcounter{equation}
\begin{equation*}}{%
\leqno(\arabic{equation})
\end{equation*}
}
\usepackage{xcolor}

\usepackage[customcolors,shade]{hf-tikz}

\newcommand{\sv}{\langle \sigma \mathit{v} \rangle}
\newcommand\CellTopTwo{\rule{0pt}{2.3ex}}

\title{Combined dark matter searches towards dwarf spheroidal galaxies with Fermi-LAT, HAWC, H.E.S.S., MAGIC, and VERITAS}

\ShortTitle{Combined dark matter searches}

\author*[a,i]{C\'eline Armand}

\author[b]{Eric Charles}
       
\author[b]{Mattia di Mauro} 
        
\author[c]{Chiara Giuri}
        
\author[d]{J. Patrick Harding}
        
\author[e]{Daniel Kerszberg}
        
 \author[f]{Tjark Miener}

\author[g]{Emmanuel Moulin}
        
\author[h]{Louise Oakes}

\author[i]{Vincent Poireau}
        
\author[c]{Elisa Pueschel}

 \author[e]{Javier Rico}
            
\author[g]{Lucia Rinchiuso}

\author[j]{Daniel Salazar-Gallegos}

\author[j]{Kirsten Tollefson}

\author[k]{Benjamin Zitzer}
        
\affiliation[a]{Astronomy Department, University of Geneva, Chemin d’Ecogia 16, 1290 Versoix, Switzerland}

\affiliation[b]{SLAC, USA}

\affiliation[c]{DESY Zeuthen, Germany}

\affiliation[d]{Los Alamos National Laboratory, USA}

\affiliation[e]{IFAE-BIST, Spain}

\affiliation[f]{IPARCOS, Universidad Complutense de Madrid, Spain}

\affiliation[g]{IRFU, CEA, Universit\'e Paris-Saclay, F-91191 Gif-sur-Yvette, France}

\affiliation[h]{Humboldt University Berlin, Germany}

\affiliation[i]{Univ. Grenoble Alpes, Univ. Savoie Mont Blanc, CNRS, LAPP, 74000 Annecy, France}

\affiliation[j]{Michigan State University, USA}

\affiliation[k]{McGill University, Canada}

\forColl{Fermi-LAT, }
\forColl{HAWC, }
\forColl{H.E.S.S.
, }
\forColl{MAGIC, }
\forColl{and VERITAS}

\emailAdd{celine.armand@lapth.cnrs.fr}
\emailAdd{echarles@slac.stanford.edu}
\emailAdd{dimauro.mattia@gmail.com}
\emailAdd{chiara.giuri@desy.de}
\emailAdd{jpharding@lanl.gov}
\emailAdd{dkerszberg@ifae.es}
\emailAdd{tmiener@ucm.es}
\emailAdd{emmanuel.moulin@cea.fr}
\emailAdd{loakes@physik.hu-berlin.de}
\emailAdd{poireau@lapp.in2p3.fr}
\emailAdd{elisa.pueschel@desy.de}
\emailAdd{jrico@ifae.es}
\emailAdd{lucia.rinchiuso@cea.fr}
\emailAdd{salaza82@msu.edu}
\emailAdd{ollefson@pa.msu.edu}
\emailAdd{bzitzer@physics.mcgill.ca}

\abstract{
Cosmological and astrophysical observations suggest that 85\% of the total matter of the Universe is made of Dark Matter (DM). 
However, its nature remains one of the most challenging and fundamental open questions of particle physics. 
Assuming particle DM, this exotic form of matter cannot consist of Standard Model (SM) particles. Many models have been developed to attempt unraveling the nature of DM such as Weakly Interacting Massive Particles (WIMPs), the most favored particle candidates.
WIMP annihilations and decay could produce SM particles which in turn hadronize and decay to give SM secondaries such as high energy $\gamma$ rays.
In the framework of indirect DM search, observations of promising targets are used to search for signatures of DM annihilation.
Among these, the dwarf spheroidal galaxies (dSphs) are commonly favored owing to their expected high DM content and negligible astrophysical background. In this work, we present the very first combination of 20 dSph observations, performed by the Fermi-LAT, HAWC, H.E.S.S., MAGIC, and VERITAS collaborations in order to maximize the sensitivity of DM searches and improve the current results. We use a joint maximum likelihood approach combining each experiment's individual analysis to derive more constraining upper limits on the WIMP DM self-annihilation cross-section as a function of DM particle mass. We present new DM constraints over the widest mass range ever reported, extending from 5 GeV to 100 TeV thanks to the combination of these five different $\gamma$-ray instruments.}

\FullConference{37$^{\rm{th}}$ International Cosmic Ray Conference (ICRC 2021)\\
		July 12th -- 23rd, 2021\\
		Online -- Berlin, Germany}

\begin{document}
\maketitle

\section{Introduction}

The nature of dark matter (DM) represents a fundamental question for the understanding of our Universe.
Observational hints at cosmological and galaxy scales such as the discrepancy between the measured rotation curves of galaxies and their theoretical predictions, the formation of large structures, and the anisotropies of the Comic Microwave Background show that DM makes up about 85\% of the total matter.

The search for DM has therefore become a priority in the scientific community where a collective effort has been made in indirect, direct, and collider searches in order to unravel its mystery. Its detection would also be a milestone in searches for Physics beyond the Standard Model (SM). 
In this talk, we focus on the indirect detection using the observations made by five gamma-ray experiments towards twenty dwarf spheroidal galaxies (dSphs).
These dSphs represent one of the most promising targets for DM indirect searches due to their high DM content and their negligible astrophysical background~\cite{PhysRevD.69.123501}.
They are all located at high Galactic latitude and no sign of very high energy emission has been detected so far in the dSphs' directions.
Gamma rays have the advantage of being neutral and do not get deflected by magnetic fields. Thus, the regions of $\gamma$-ray production can be traced back from the incident direction. The observations are therefore performed based on this property where the telescopes are directly pointing to the sources.
%

This work represents a collective effort between three imaging atmospheric Cherenkov telescope (IACT) arrays  H.E.S.S., MAGIC, and VERITAS, the water Cherenkov array HAWC, and the space-borne telescope {\it Fermi}-LAT, which agreed on sharing their data previously published individually. 
The goal of our study is to combine the individual upper limits published by each collaboration in order to optimize the statistics and increase the sensitivity to potential DM signals. The combination brings the novelty of extending the upper and lower boundaries of the energy range and the derivation of the upper limits on the DM annihilation cross-section over the widest DM particle mass range ever.
%
In this work, each of the five collaborations performed the analysis of their own data sets using a common DM model to optimize the data handling at different energy, angular resolutions, and sensitive energy ranges of the various instruments.
By following this procedure, we also avoid the need for sharing raw data and instrument response functions (IRFs) outside the collaborations.
As no significant excess was detected from the selected sources, nor in their combination, we derive upper limits on the DM annihilation cross-section as a function of the DM particle mass by combining the likelihood functions of all dSphs and all experiments.

\section{Experiments} \label{Sec:Experiments}

\subsection{Fermi-LAT}

The {\it Fermi}-Large Area Telescope ({\it Fermi-LAT}) is the collaboration which operates the pair conversion Large Area Telescope (LAT) carried by the Fermi satellite orbiting the Earth at an altitude of 565 km since 2008. The telescope has a wide field of view covering about 20\% of the sky and scans the whole sky every 3 hours in the energy range between $\sim 20$ MeV and 1~TeV. {\it Fermi}-LAT thus covers the lowest energy region of this study. Detailed descriptions of the detector and its performance can be found in \cite{2012ApJS..203....4A}.

\subsection{HAWC}

The High-Altitude Water Cherenkov (HAWC) detector is a high-energy $\gamma$-ray telescope located at Sierra Negra, Mexico at 4100~m altitude and consists of an array consisting of 300 water Cherenkov detectors (WCD) covering an area of 22,000~$\rm{m}^2$. The WCD are sensitive to $\gamma$-ray events of energies ranging from 300 GeV to a couple hundred TeV \cite{abeysekara}. The experiment covers a field of view of 15\% of the sky at all times.

\subsection{H.E.S.S.}

The High Energy Stereosocpic System (H.E.S.S.) experiment is an array consisting of five IACTs designed to detect brief and faint flashes of Cherenkov radiation generated by very high energy $\gamma$ rays between $\sim$30 GeV and $\sim 100$ TeV. The telescope array is located in central Namibia in the Khomas Highland region at 1,800 m above sea level~\cite{Aharonian:2006pe} at 110 km south west of Windhoek. The four small telescopes are equipped with a 12~m reflector while the central one is 28~m. The array collects the $\gamma$ rays within a field of view of 5$^\circ$. 

\subsection{MAGIC}
The Major Atmospheric Gamma-ray Imaging Cherenkov (MAGIC) telescope array consists of two telescopes of 17 m diameter reflector situated at the Roque de los Muchachos Observatory on the Canary Island of La Palma, Spain, 2,200~m above sea level. MAGIC is sensitive to very high energy $\gamma$-ray events above $\sim 50$ GeV~\cite{Sitarek} and is equipped with fast imaging cameras with a field of view of 3.5\degree.

\subsection{VERITAS}
The Very Energetic Radiation Imaging Telescope Array System (VERITAS) is an array of four telescopes of 12~m reflector located at the Fred Lawrence Whipple Observatory in Southern Arizona. The telescope array is sensitive to a very high energetic band from $\sim$ 85~GeV up to $\sim 30$~TeV whose events are recorded within a $3.5\degree$ field of view~\cite{Park}.

\section{DM signal} \label{sec:DM}

The differential flux of $\gamma$ rays from the self-annihilation of Majorana DM particles is given by:
\begin{equation}
    \label{eq:DM_flux}
    \frac{\text{d}^2\Phi \left(\langle \sigma v \rangle, J \right)}{\text{d}E \text{d}\Omega} = \frac{1}{4\pi} \frac{\langle \sigma v \rangle}{2m^{2}_{\chi}} \sum_{f} \text{BR}_{f}\frac{\text{d}N_{f}}{\text{d}E} \times \frac{ \text{d}J}{\text{d}\Omega}.
\end{equation}

The first term contains the mass $m_{\chi}$ of the DM particles in GeV and their annihilation cross-section averaged over the velocity distribution $\langle \sigma v \rangle$ in $\rm{cm}^3\rm{s}^{-1}$. It also carries the differential spectrum $dN_f /dE$ for a given annihilation channel $f$.
Since we do not assume any specific particle physics model, each channel is treated individually where the branching ratio $\text{BR}_{f} = 100\%$.
The second term, known as the astrophysical $J$ factor, describes the amount of DM annihilations within a source or a region of the sky. 

The differential $J$-factor is defined as the integral of the square of the DM density distribution $\rho_{\rm{DM}}$ along the line-of-sight (l.o.s.):
\begin{equation}
  \frac{\text{d}J}{\text{d}\Omega} = \displaystyle{\int_\mathrm{l.o.s.} \: \rho_{\textrm{DM}}^2(r(s,\theta)) \:ds },
  \label{dwarf_J}
\end{equation}
where $\rho_{\textrm{DM}}$ is assumed to be spherically symmetric for all considered dSphs and depends on the distance to the center of the source $r$. This distance can also be expressed in terms of the distance $s$ from Earth along the line of sight, and the angular distance $\theta$ with respect to the center of the source, as $r(s, \theta) = (s^2 +d^2 -2sd\cos\theta)^{1/2}$, where $d$ is the distance between the Earth and the source. 
The $J$ factor computation is usually performed through a Jeans analysis based on the spherical Jeans equations~\cite{Bonnivard:2014kza, Geringer-Sameth:2014yza, binney2011galactic, Bonnivard:2015xpq}.
This technique relies on the spectroscopic data and assumes that the galaxies are in steady-state hydrodynamic equilibrium, have a spherical symmetry, and are non-rotating systems to reconstruct the galactic dynamics.
In this work, we use the $J$ factors produced by Geringer-Sameth \emph{et al.}~\cite{Geringer-Sameth:2014yza}.

\section{Joint likelihood analysis}\label{Sec:Likelihood}

\subsection{Dataset}

Twenty classical and ultrafaint dwarf spheroidal galaxies are selected for the combination. All were observed by one or more instruments and previously published by individual collaborations. Table~\ref{tab:dsphs_list} presents the list of dwarf galaxies used this project and the experiments with which they were observed.

\begin{table}[h!]
\small{
\centering{
\begin{tabular}{cccc}
\hline
\hline
Source name          & Experiments  & Distance & $\log_{10}J$           \\
 & & \scriptsize{(kpc)} &  \scriptsize{$\log_{10}(\rm{GeV}^2 \rm{cm}^{-5}\rm{sr})$}   \\
\hline
\CellTopTwo{}
Bootes I             & {\it Fermi}-LAT, HAWC, VERITAS &$66$ & $18.24^{+0.40}_{-0.37}$  \\
\CellTopTwo{}
Canes Venatici I     & {\it Fermi}-LAT   &$218$ & $17.44^{+0.37}_{-0.28}$           \\
\CellTopTwo{}
Canes Venatici II     & {\it Fermi}-LAT, HAWC & $160$& $17.65^{+0.45}_{-0.43}$    \\
\CellTopTwo{}
Carina     & {\it Fermi}-LAT, H.E.S.S. & $105$&  $17.92^{+0.19}_{-0.11}$  \\
\CellTopTwo{}
Coma Berenices     & {\it Fermi}-LAT, HAWC, H.E.S.S., MAGIC & $44$&  $19.02^{+0.37}_{-0.41}$   \\
\CellTopTwo{}
Draco     & {\it Fermi}-LAT, HAWC, MAGIC, VERITAS & $76$ &  $19.05^{+0.22}_{-0.21}$  \\
\CellTopTwo{}
Fornax    & {\it Fermi}-LAT, H.E.S.S.   & $147$ & $17.84^{+0.11}_{-0.06}$  \\
\CellTopTwo{}
Hercules   & {\it Fermi}-LAT, HAWC  & $132$ & $16.86^{+0.74}_{-0.68}$  \\
\CellTopTwo{}
Leo I     & {\it Fermi}-LAT, HAWC  & $254$ & $17.84^{+0.20}_{-0.16}$  \\
\CellTopTwo{}
Leo II     & {\it Fermi}-LAT, HAWC  & $233$ & $17.97^{+0.20}_{-0.18}$   \\
\CellTopTwo{}
Leo IV   & {\it Fermi}-LAT, HAWC  & $154$&  $16.32^{+1.06}_{-1.70}$  \\
\CellTopTwo{}
Leo T     & {\it Fermi}-LAT  &$417$& $17.11^{+0.44}_{-0.39}$    \\
\CellTopTwo{}
Leo V     & {\it Fermi}-LAT & $178$&  $16.37^{+0.94}_{-0.87}$  \\
\CellTopTwo{}
Sculptor     & {\it Fermi}-LAT, H.E.S.S.  & $86$& $18.57^{+0.07}_{-0.05}$  \\
\CellTopTwo{}
Segue I     & {\it Fermi}-LAT, HAWC, MAGIC, VERITAS   & $23$ &  $19.36^{+0.32}_{-0.35}$  \\
\CellTopTwo{}
Segue II     & {\it Fermi}-LAT   & $35$ & $16.21^{+1.06}_{-0.98}$ \\
\CellTopTwo{}
Sextans    & {\it Fermi}-LAT, HAWC &$86$ & $17.92^{+0.35}_{-0.29}$  \\
\CellTopTwo{}
Ursa Major I     & {\it Fermi}-LAT, HAWC  & $97$ & $17.87^{+0.56}_{-0.33}$  \\
\CellTopTwo{}
Ursa Major II     & {\it Fermi}-LAT, HAWC, MAGIC  &$32$& $19.42^{+0.44}_{-0.42}$   \\
\CellTopTwo{}
Ursa Minor     & {\it Fermi}-LAT, VERITAS  & $76$ & $18.95^{+0.26}_{-0.18}$  \\
\hline
\hline
\end{tabular}
}
\caption{Summary of the relevant properties of the dSphs included in the combination of {\it Fermi}-LAT, HAWC, H.E.S.S., MAGIC, and VERITAS likelihood functions. The list of the observed dwarf galaxies is presented in column 1 with the instruments that performed the observations in column 2. Their heliocentric distance and  $J$ factor with their estimated $\pm 1\sigma$ uncertainties are listed in columns 3 and 4 respectively. The $J$ factors are given for a source extension truncated at the outermost observed star with their estimated $\pm 1\sigma$ uncertainties. }
\label{tab:dsphs_list}}
\end{table}

\subsection{Combination principle}

Our search for DM is carried out using a technique of maximum likelihood in which the profile likelihood ratio $\lambda$ is a function of the annihilation cross-section $\langle \sigma v \rangle$, \emph{i.e.} the parameter of interest, and reads as:

\begin{equation}
    \label{Eq:likelihood_profile}
\lambda \left( \sv \mid \boldsymbol{\mathcal{D}_{\text{dSphs}}} \right) = \frac{\mathcal{L} \left( \sv ; \boldsymbol{\hat{\hat{\nu}}} \mid \boldsymbol{\mathcal{D}_{\text{dSphs}}} \right)}{\mathcal{L} \left( \widehat{\langle\sigma v \rangle} ; \boldsymbol{\hat{\nu}} \mid \boldsymbol{\mathcal{D}_{\text{dSphs}}} \right)} ,
\end{equation}
where $\boldsymbol{\mathcal{D}_{\text{dSphs}}}$ is the dataset, $\boldsymbol{\nu}$ represents the nuisance parameters, $\widehat{\langle\sigma v \rangle}$ and $\boldsymbol{\hat{\nu}}$ are the values that maximize $\mathcal{L}$ globally, and $\boldsymbol{\hat{\hat{\nu}}}$ the values that maximize $\mathcal{L}$ for a given value of $\sv$. 
The joint likelihood function $\mathcal{L}$ describing all measurements is the product of the individual likelihood functions of all instruments and all dSphs and is given by:
\begin{equation}
    \label{Eq:dSph_combination}
    \mathcal{L} \left( \sv ; \boldsymbol{\nu} \mid \boldsymbol{\mathcal{D}_{\text{dSphs}}} \right) = \prod_{l=1}^{N_{\text{dSphs}}} \mathcal{L}_{\text{dSph},l} \left( \sv ; J_{l}, \boldsymbol{\nu_{l}} \mid \boldsymbol{\mathcal{D}_{l,\text{measured}}} \right) \times \mathcal{J}_l \left( J_{l} \mid J_{l,\text{obs}}, \sigma_{\log{J_l}} \right).
\end{equation}
The quantity $ N_{\text{dSphs}} = 20 $ is the total number of dSphs; $ \boldsymbol{\mathcal{D}_{l,\text{measured}}} $ is the dataset from gamma-ray observations for the $l$-th dSph; $ \boldsymbol{\nu_{l}} $ is the set of nuisance parameters associated to the $l$-th dSph, excluding $ J_{l} $; and $ J_{l} $ is the total $J$ factor of the $l$-th dSph, whose value can be found in Tab.~\ref{tab:dsphs_list}
; $\log_{10} J_{l,\text{obs}}$ and $ \sigma_{\log{J_l}} $ are obtained from the fit (see Jeans analysis in Sec.~\ref{sec:DM}) of a log-normal function of $J_{l,\text{obs}}$ to the posterior distribution of $ J_{l} $ \cite{2015PhRvL.115w1301A}. 

\subsection{Shared data format}
In order to perform the combination of the observations, a table of test statistic (TS) values is provided by each experiment for the annihilation channels, $b\bar{b}$ and $\tau^+\tau^-$, for each set of $m_{\chi}$ and $\langle \sigma v \rangle$. All collaborations agreed on 63 DM masses ranging from 10~GeV to 100~TeV for all continuum channels following the mass spacing of~\cite{Cirelli} to avoid an interpolation. The $\langle \sigma v \rangle$ range is defined between $10^{-28}~\rm{cm}^3\rm{s}^{-1}$ and $10^{-18}~\rm{cm}^3\rm{s}^{-1}$ and is logarithmically spaced in 1001 values. 

\subsection{Statistical uncertainty bands}
The $68\%$ (1$\sigma$) and $95\%$ (2$\sigma$) containment bands are derived by individual experiments by performing 300 Poisson realizations of the background events. Each collaboration provides the results of their statistical uncertainties in the same format as for the nominal values which are then combined following the same procedure as the combination of the nominal upper limits.

\section{Results and discussion}\label{Sec:Results}
\begin{figure}[h!]
\includegraphics[width=0.93\textwidth]{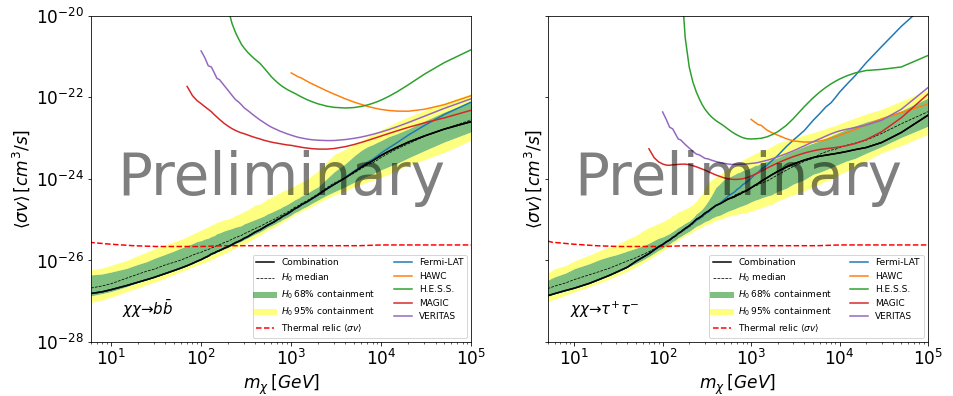}
\vspace{-0.1cm}

\caption{Upper limits at 95\% confidence level on $\sv$ as a function of the DM mass for the annihilation channels $b\bar{b}$ (left) and $\tau^+ \tau^-$ (right), using the set of $J$ factors from Ref.~\cite{Geringer-Sameth:2014yza}. The black solid line represents the observed combined limit, the black dashed line is the median of the null hypothesis corresponding to the expected limit, while the green and yellow bands show the 68\% and 95\% containment bands. Combined upper limits for each individual detector are also indicated as solid, colored lines. }

\label{fig:bb_tautau_UL}
\end{figure}

No significant DM signal has been observed by any of the five instruments.
We therefore present the results of the combined upper limits at 95\% C.L. on the DM annihilation cross-section $\langle \sigma v \rangle$ in the case of two annihilation channels, $b\bar{b}$ and $\tau^+ \tau^-$, using all the data collected towards the twenty dSphs. We note that we selected these hadronic and leptonic channels as the follow up of our previous results presented at ICRC 2019~\cite{Oakes:2019ywx}.
We set our upper limits by solving $\mathrm{TS} = -2 \ln{ \lambda(\sv)}$ for $\sv$, with $\mathrm{TS}= 2.71 $. The value 2.71 represents the 95\% confidence level of a one-sided distribution assuming the test statistics behaves like a $\chi^2$ distribution with one degree of freedom. The combination is performed using two independent public analysis software packages, \texttt{gLike} \cite{javier_rico_2021_4601451} and \texttt{LklCombiner} \cite{tjark_miener_2021_4597500}, that provide compatible results.
The combined upper limits are presented in Fig.~\ref{fig:bb_tautau_UL} and are given with their 68\% ($1\sigma$) and 95\% ($2\sigma$) containment bands. These limits (solid black lines) are expected to be close to the median limit (dashed black lines) as no signal is present. 
We obtain upper limits within the 2 $\sigma$ expected bands for the two annihilation channels $b\bar{b}$ and $\tau^+ \tau^-$.
The individual limits produced by each experiment are also indicated in the figures as a comparison to our new combined results.
Below \textasciitilde500~GeV, the DM limits are largely dominated by the {\it Fermi}-LAT experiment.
Between \textasciitilde500 GeV to \textasciitilde10 TeV, {\it Fermi}-LAT continues to dominate for the hadronic DM channel then above \textasciitilde10 TeV, the IACTs (H.E.S.S., MAGIC, and VERITAS) and HAWC take over.
In the case of the leptonic channel, both the IACTs and HAWC contribute significantly to the DM limit from \textasciitilde1 TeV to \textasciitilde100 TeV.

\vspace{-0.2cm}

We observe that the combined DM constraints from all five telescopes are 2 to 3 times stronger than any individual telescope for multi-TeV DM.
The selection of multiple targets increases statistics used to probe these sources and allows us to derive upper limits spanning the largest mass range of any WIMP DM search. We note that these limits depend on the choice of the annihilation channels and are driven by the objects with the highest $J$ factors that can be observed. The ultrafaint dSphs, containing a few tens of bright stars only, can be subject to large systematic uncertainties for the determination of their $J$-factors such as Segue I.
The derivation of upper limits through 6 additional annihilation channels is currently in progress, with 5 other continuum channels and the monoenergetic channel $\gamma \gamma$. A further analysis using a second $J$ factor set derived by~\cite{Bonnivard:2014kza, Bonnivard:2015xpq} is also yet to come in order to study the systematics induced by the choice of $J$ factor.



{\small
\bibliographystyle{ieeetr}
\bibliography{References}

\begin{thebibliography}{10}

\bibitem{PhysRevD.69.123501}
N.~W. Evans, F.~Ferrer, and S.~Sarkar, ``A travel guide to the dark matter
  annihilation signal,'' {\em Phys. Rev. D}, vol.~69, p.~123501, Jun 2004.

\bibitem{2012ApJS..203....4A}
M.~{Ackermann} {\em et~al.}, ``{The {\it Fermi} Large Area Telescope on Orbit:
  Event Classification, Instrument Response Functions, and Calibration},'' {\em
  Astrophysical Journal Supplement Series}, vol.~203, p.~4, Nov. 2012.

\bibitem{abeysekara}
A.~U. Abeysekara {\em et~al.}, ``{Observation of the Crab Nebula with the HAWC
  Gamma-Ray Observatory},'' {\em Astrophys. J.}, vol.~843, no.~1, p.~39, 2017.

\bibitem{Aharonian:2006pe}
F.~Aharonian {\em et~al.}, ``{Observations of the Crab Nebula with H.E.S.S},''
  {\em Astron. Astrophys.}, vol.~457, pp.~899--915, 2006.

\bibitem{Sitarek}
J.~Sitarek {\em et~al.}, ``{Physics performance of the upgraded MAGIC
  telescopes obtained with Crab Nebula data},'' in {\em {Proceedings, 33rd
  International Cosmic Ray Conference (ICRC2013): Rio de Janeiro, Brazil, July
  2-9, 2013}}, p.~0074, 2013.

\bibitem{Park}
N.~Park, ``{Performance of the VERITAS experiment},'' in {\em {Proceedings,
  34th International Cosmic Ray Conference (ICRC2015): The Hague, The
  Netherlands, 30th July - 6th August}}, vol.~34, p.~771, 2015.

\bibitem{Bonnivard:2014kza}
V.~Bonnivard, C.~Combet, D.~Maurin, and M.~G. Walker, ``{Spherical Jeans
  analysis for dark matter indirect detection in dwarf spheroidal galaxies -
  Impact of physical parameters and triaxiality},'' {\em Mon. Not. Roy. Astron.
  Soc.}, vol.~446, pp.~3002--3021, 2015.

\bibitem{Geringer-Sameth:2014yza}
A.~Geringer-Sameth, S.~M. Koushiappas, and M.~Walker, ``{Dwarf galaxy
  annihilation and decay emission profiles for dark matter experiments},'' {\em
  Astrophys. J.}, vol.~801, no.~2, p.~74, 2015.

\bibitem{binney2011galactic}
J.~Binney and S.~Tremaine, {\em Galactic Dynamics: Second Edition}.
\newblock Princeton Series in Astrophysics, Princeton University Press, 2011.

\bibitem{Bonnivard:2015xpq}
V.~Bonnivard {\em et~al.}, ``{Dark matter annihilation and decay in dwarf
  spheroidal galaxies: The classical and ultrafaint dSphs},'' {\em Mon. Not.
  Roy. Astron. Soc.}, vol.~453, no.~1, pp.~849--867, 2015.

\bibitem{2015PhRvL.115w1301A}
M.~{Ackermann} {\em et~al.}, ``{Searching for Dark Matter Annihilation from
  Milky Way Dwarf Spheroidal Galaxies with Six Years of {\it Fermi} Large Area
  Telescope Data},'' {\em Physical Review Letters}, vol.~115, p.~231301, Dec.
  2015.

\bibitem{Cirelli}
M.~Cirelli, ``{PPPC 4 DM ID: a poor particle physicist cookbook for dark matter
  indirect detection},'' {\em {Journal of Cosmology and Astroparticle
  Physics}}, vol.~2011, 2011.

\bibitem{Oakes:2019ywx}
L.~Oakes {\em et~al.}, ``{Combined Dark Matter searches towards dwarf
  spheroidal galaxies with $Fermi$-LAT, HAWC, HESS, MAGIC and VERITAS},'' {\em
  PoS}, vol.~ICRC2019, p.~012, 2021.

\bibitem{javier_rico_2021_4601451}
J.~Rico, C.~Nigro, D.~Kerszberg, T.~Miener, and J.~Aleksic, ``{gLike: numerical
  maximization of heterogeneous joint likelihood functions of a common free
  parameter plus nuisance parameters}.''
  https://doi.org/10.5281/zenodo.4601451, Mar. 2021.

\bibitem{tjark_miener_2021_4597500}
T.~Miener and D.~Nieto, ``{LklCom: Combining likelihoods from different
  experiments}.'' https://doi.org/10.5281/zenodo.4597500, Mar. 2021.

\end{thebibliography}
}

\clearpage
\section*{Full Author List}

\subsection*{{\it Fermi}-LAT}

The \textit{Fermi}-LAT Collaboration acknowledges support for LAT development, operation and data analysis from NASA and DOE (United States), CEA/Irfu and IN2P3/CNRS (France), ASI and INFN (Italy), MEXT, KEK, and JAXA (Japan), and the K.A.~Wallenberg Foundation, the Swedish Research Council and the National Space Board (Sweden). Science analysis support in the operations phase from INAF (Italy) and CNES (France) is also gratefully acknowledged. This work performed in part under DOE Contract DE-AC02-76SF00515.

\subsection*{HAWC}

\noindent
A.U. Abeysekara$^{48}$,
A. Albert$^{21}$,
R. Alfaro$^{14}$,
C. Alvarez$^{41}$,
J.D. Álvarez$^{40}$,
J.R. Angeles Camacho$^{14}$,
J.C. Arteaga-Velázquez$^{40}$,
K. P. Arunbabu$^{17}$,
D. Avila Rojas$^{14}$,
H.A. Ayala Solares$^{28}$
% ...

\noindent
{\scriptsize
% ...
$^{14}$Instituto de Física, Universidad Nacional Autónoma de México, Ciudad de México, México,
% ...
$^{17}$Instituto de Geofísica, Universidad Nacional Autónoma de México, Ciudad de México, México,
% ...
$^{21}$Physics Division, Los Alamos National Laboratory, Los Alamos, NM, USA,
% ...
$^{28}$Department of Physics, Pennsylvania State University, University Park, PA, USA,
% ...
$^{40}$Instituto de Física y Matemáticas, Universidad Michoacana de San Nicolás de Hidalgo, Morelia, Michoacán, México,
$^{41}$FCFM-MCTP, Universidad Autónoma de Chiapas, Tuxtla Gutiérrez, Chiapas, México,
% ...
$^{48}$Department of Physics and Astronomy, University of Utah, Salt Lake City, UT, USA,
% ...
}

\vspace{0.3cm}

We acknowledge the support from: the US National Science Foundation (NSF); the US Department of Energy Office of High-Energy Physics; the Laboratory Directed Research and Development (LDRD) program of Los Alamos National Laboratory; Consejo Nacional de Ciencia y Tecnolog\'ia (CONACyT), M\'exico, grants 271051, 232656, 260378, 179588, 254964, 258865, 243290, 132197, A1-S-46288, A1-S-22784, c\'atedras 873, 1563, 341, 323, Red HAWC, M\'exico; DGAPA-UNAM grants IG101320, IN111716-3, IN111419, IA102019, IN110621, IN110521; VIEP-BUAP; PIFI 2012, 2013, PROFOCIE 2014, 2015; the University of Wisconsin Alumni Research Foundation; the Institute of Geophysics, Planetary Physics, and Signatures at Los Alamos National Laboratory; Polish Science Centre grant, DEC-2017/27/B/ST9/02272; Coordinaci\'on de la Investigaci\'on Cient\'ifica de la Universidad Michoacana; Royal Society - Newton Advanced Fellowship 180385; Generalitat Valenciana, grant CIDEGENT/2018/034; Chulalongkorn University’s CUniverse (CUAASC) grant; Coordinaci\'on General Acad\'emica e Innovaci\'on (CGAI-UdeG), PRODEP-SEP UDG-CA-499; Institute of Cosmic Ray Research (ICRR), University of Tokyo, H.F. acknowledges support by NASA under award number 80GSFC21M0002. We also acknowledge the significant contributions over many years of Stefan Westerhoff, Gaurang Yodh and Arnulfo Zepeda Dominguez, all deceased members of the HAWC collaboration. Thanks to Scott Delay, Luciano D\'iaz and Eduardo Murrieta for technical support.

\subsection*{H.E.S.S.}

\noindent
H.~Abdallah$^{1}$,
R.~Adam$^{2}$,
F.~Aharonian$^{3,4,5}$,
F.~Ait Benkhali$^{3}$,
E.O.~Ang{\"u}ner$^{6}$,
C.~Arcaro$^{1}$,
C.~Armand$^{7,44}$,
T.~Armstrong$^{8}$,
H.~Ashkar$^{9}$,
M.~Backes$^{10}$
% ...
\\
\noindent
{\scriptsize
$^{1}${Centre for Space Research, North-West University, Potchefstroom 2520, South Africa}
$^{2}${Laboratoire Leprince-Ringuet, École Polytechnique, CNRS, Institut Polytechnique de Paris, F-91128 Palaiseau, France}
$^{3}${Max-Planck-Institut f\"ur Kernphysik, P.O. Box 103980, D 69029 Heidelberg, Germany}
$^{4}${Dublin Institute for Advanced Studies, 31 Fitzwilliam Place, Dublin 2, Ireland}
$^{5}${High Energy Astrophysics Laboratory, RAU, 123 Hovsep Emin St Yerevan 0051, Armenia}
$^{6}${Aix Marseille Universit\'e, CNRS/IN2P3, CPPM, Marseille, France}
$^{7}${Laboratoire d'Annecy de Physique des Particules, Univ. Grenoble Alpes, Univ. Savoie Mont Blanc, CNRS, LAPP, 74000 Annecy, France}
$^{8}${University of Oxford, Department of Physics, Denys Wilkinson Building, Keble Road, Oxford OX1 3RH, UK}
$^{9}${IRFU, CEA, Universit\'e Paris-Saclay, F-91191 Gif-sur-Yvette, France}
$^{10}${University of Namibia, Department of Physics, Private Bag 13301, Windhoek, Namibia}
% ...
$^{44}${Astronomy Department, University of Geneva, Chemin d’Ecogia 16, 1290 Versoix, Switzerland}
}

\vspace{0.3cm}

The support of the Namibian authorities and of the University of Namibia in facilitating 
the construction and operation of H.E.S.S. is gratefully acknowledged, as is the support 
by the German Ministry for Education and Research (BMBF), the Max Planck Society, the 
German Research Foundation (DFG), the Helmholtz Association, the Alexander von Humboldt Foundation, 
the French Ministry of Higher Education, Research and Innovation, the Centre National de la 
Recherche Scientifique (CNRS/IN2P3 and CNRS/INSU), the Commissariat à l’énergie atomique 
et aux énergies alternatives (CEA), the U.K. Science and Technology Facilities Council (STFC), 
the Knut and Alice Wallenberg Foundation, the National Science Centre, Poland grant no. 2016/22/M/ST9/00382, 
the South African Department of Science and Technology and National Research Foundation, the 
University of Namibia, the National Commission on Research, Science \& Technology of Namibia (NCRST), 
the Austrian Federal Ministry of Education, Science and Research and the Austrian Science Fund (FWF), 
the Australian Research Council (ARC), the Japan Society for the Promotion of Science and by the 
University of Amsterdam. We appreciate the excellent work of the technical support staff in Berlin, 
Zeuthen, Heidelberg, Palaiseau, Paris, Saclay, Tübingen and in Namibia in the construction and 
operation of the equipment. This work benefitted from services provided by the H.E.S.S. 
Virtual Organisation, supported by the national resource providers of the EGI Federation. 

\subsection*{MAGIC}

\noindent
V.~A.~Acciari$^{1}$,
S.~Ansoldi$^{2,41}$,
L.~A.~Antonelli$^{3}$,
A.~Arbet Engels$^{4}$,
M.~Artero$^{5}$,
K.~Asano$^{6}$,
D.~Baack$^{7}$,
A.~Babi\'c$^{8}$,
A.~Baquero$^{9}$,
U.~Barres de Almeida$^{10}$
% ...
\\
\noindent
{\scriptsize
$^{1}$ {Instituto de Astrof\'isica de Canarias and Dpto. de  Astrof\'isica, Universidad de La Laguna, E-38200, La Laguna, Tenerife, Spain}
$^{2}$ {Universit\`a di Udine and INFN Trieste, I-33100 Udine, Italy}
$^{3}$ {National Institute for Astrophysics (INAF), I-00136 Rome, Italy}
$^{4}$ {ETH Z\"urich, CH-8093 Z\"urich, Switzerland}
$^{5}$ {Institut de F\'isica d'Altes Energies (IFAE), The Barcelona Institute of Science and Technology (BIST), E-08193 Bellaterra (Barcelona), Spain}
$^{6}$ {Japanese MAGIC Group: Institute for Cosmic Ray Research (ICRR), The University of Tokyo, Kashiwa, 277-8582 Chiba, Japan}
$^{7}$ {Technische Universit\"at Dortmund, D-44221 Dortmund, Germany}
$^{8}$ {Croatian MAGIC Group: University of Zagreb, Faculty of Electrical Engineering and Computing (FER), 10000 Zagreb, Croatia}
$^{9}$ {IPARCOS Institute and EMFTEL Department, Universidad Complutense de Madrid, E-28040 Madrid, Spain}
$^{10}$ {Centro Brasileiro de Pesquisas F\'isicas (CBPF), 22290-180 URCA, Rio de Janeiro (RJ), Brazil}
% ...
$^{41}$ {also at International Center for Relativistic Astrophysics (ICRA), Rome, Italy}
% ...
}

\vspace{0.3cm}

We would like to thank the Instituto de Astrofísica de Canarias for the excellent working conditions at the Observatorio del Roque de los Muchachos in La Palma. The financial support of the German BMBF, MPG and HGF; the Italian INFN and INAF; the Swiss National Fund SNF; the ERDF under the Spanish Ministerio de Ciencia e Innovación (MICINN) (PID2019-104114RB-C31, PID2019-104114RB-C32, PID2019-104114RB-C33, PID2019-105510GB-C31,PID2019-107847RB-C41, PID2019-107847RB-C42, PID2019-107988GB-C22); the Indian Department of Atomic Energy; the Japanese ICRR, the University of Tokyo, JSPS, and MEXT; the Bulgarian Ministry of Education and Science, National RI Roadmap Project DO1-400/18.12.2020 and the Academy of Finland grant nr. 320045 is gratefully acknowledged. This work was also supported by the Spanish Centro de Excelencia "Severo Ochoa" (SEV-2016-0588, CEX2019-000920-S), the Unidad de Excelencia "María de Maeztu" (CEX2019-000918-M, MDM-2015-0509-18-2) and by the CERCA program of the Generalitat de Catalunya; by the Croatian Science Foundation (HrZZ) Project IP-2016-06-9782 and the University of Rijeka Project 13.12.1.3.02; by the DFG Collaborative Research Centers SFB823/C4 and SFB876/C3; the Polish National Research Centre grant UMO-2016/22/M/ST9/00382; and by the Brazilian MCTIC, CNPq and FAPERJ.

\subsection*{VERITAS}

\noindent
C.~B.~Adams$^{1}$,
A.~Archer$^{2}$,
W.~Benbow$^{3}$,
A.~Brill$^{1}$,
J.~H.~Buckley$^{4}$,
M.~Capasso$^{5}$,
J.~L.~Christiansen$^{6}$,
A.~J.~Chromey$^{7}$, 
M.~Errando$^{4}$,
A.~Falcone$^{8}$,
K.~A.~Farrell$^{9}$,
Q.~Feng$^{5}$,
G.~M.~Foote$^{10}$,
L.~Fortson$^{11}$,
A.~Furniss$^{12}$,
A.~Gent$^{13}$,
G.~H.~Gillanders$^{14}$,
C.~Giuri$^{15}$,
O.~Gueta$^{15}$,
D.~Hanna$^{16}$,
O.~Hervet$^{17}$,
J.~Holder$^{10}$,
B.~Hona$^{18}$,
T.~B.~Humensky$^{1}$,
W.~Jin$^{19}$,
P.~Kaaret$^{20}$,
M.~Kertzman$^{2}$,
T.~K.~Kleiner$^{15}$,
S.~Kumar$^{16}$,
M.~J.~Lang$^{14}$,
M.~Lundy$^{16}$,
G.~Maier$^{15}$,
C.~E~McGrath$^{9}$,
P.~Moriarty$^{14}$,
R.~Mukherjee$^{5}$,
D.~Nieto$^{21}$,
M.~Nievas-Rosillo$^{15}$,
S.~O'Brien$^{16}$,
R.~A.~Ong$^{22}$,
A.~N.~Otte$^{13}$,
S.~R. Patel$^{15}$,
S.~Patel$^{20}$,
K.~Pfrang$^{15}$,
M.~Pohl$^{23,15}$,
R.~R.~Prado$^{15}$,
E.~Pueschel$^{15}$,
J.~Quinn$^{9}$,
K.~Ragan$^{16}$,
P.~T.~Reynolds$^{24}$,
D.~Ribeiro$^{1}$,
E.~Roache$^{3}$,
J.~L.~Ryan$^{22}$,
I.~Sadeh$^{15}$,
M.~Santander$^{19}$,
G.~H.~Sembroski$^{25}$,
R.~Shang$^{22}$,
D.~Tak$^{15}$,
V.~V.~Vassiliev$^{22}$,
A.~Weinstein$^{7}$,
D.~A.~Williams$^{17}$,
and 
T.~J.~Williamson$^{10}$\\
\noindent

{\scriptsize
$^1${Physics Department, Columbia University, New York, NY 10027, USA}
$^{2}${Department of Physics and Astronomy, DePauw University, Greencastle, IN 46135-0037, USA}
$^3${Center for Astrophysics $|$ Harvard \& Smithsonian, Cambridge, MA 02138, USA}
$^4${Department of Physics, Washington University, St. Louis, MO 63130, USA}
$^5${Department of Physics and Astronomy, Barnard College, Columbia University, NY 10027, USA}
$^6${Physics Department, California Polytechnic State University, San Luis Obispo, CA 94307, USA} 
$^7${Department of Physics and Astronomy, Iowa State University, Ames, IA 50011, USA}
$^8${Department of Astronomy and Astrophysics, 525 Davey Lab, Pennsylvania State University, University Park, PA 16802, USA}
$^9${School of Physics, University College Dublin, Belfield, Dublin 4, Ireland}
$^{10}${Department of Physics and Astronomy and the Bartol Research Institute, University of Delaware, Newark, DE 19716, USA}
$^{11}${School of Physics and Astronomy, University of Minnesota, Minneapolis, MN 55455, USA}
$^{12}${Department of Physics, California State University - East Bay, Hayward, CA 94542, USA}
$^{13}${School of Physics and Center for Relativistic Astrophysics, Georgia Institute of Technology, 837 State Street NW, Atlanta, GA 30332-0430}
$^{14}${School of Physics, National University of Ireland Galway, University Road, Galway, Ireland}
$^{15}${DESY, Platanenallee 6, 15738 Zeuthen, Germany}
$^{16}${Physics Department, McGill University, Montreal, QC H3A 2T8, Canada}
$^{17}${Santa Cruz Institute for Particle Physics and Department of Physics, University of California, Santa Cruz, CA 95064, USA}
$^{18}${Department of Physics and Astronomy, University of Utah, Salt Lake City, UT 84112, USA}
$^{19}${Department of Physics and Astronomy, University of Alabama, Tuscaloosa, AL 35487, USA}
$^{20}${Department of Physics and Astronomy, University of Iowa, Van Allen Hall, Iowa City, IA 52242, USA}
$^{21}${Institute of Particle and Cosmos Physics, Universidad Complutense de Madrid, 28040 Madrid, Spain}
$^{22}${Department of Physics and Astronomy, University of California, Los Angeles, CA 90095, USA}
$^{23}${Institute of Physics and Astronomy, University of Potsdam, 14476 Potsdam-Golm, Germany}
$^{24}${Department of Physical Sciences, Munster Technological University, Bishopstown, Cork, T12 P928, Ireland}
$^{25}${Department of Physics and Astronomy, Purdue University, West Lafayette, IN 47907, USA}
}

\vspace{0.3cm}

This research is supported by grants from the U.S. Department of Energy Office of Science, the U.S. National Science Foundation and the Smithsonian Institution, by NSERC in Canada, and by the Helmholtz Association in Germany. This research used resources provided by the Open Science Grid, which is supported by the National Science Foundation and the U.S. Department of Energy's Office of Science, and resources of the National Energy Research Scientific Computing Center (NERSC), a U.S. Department of Energy Office of Science User Facility operated under Contract No. DE-AC02-05CH11231. We acknowledge the excellent work of the technical support staff at the Fred Lawrence Whipple Observatory and at the collaborating institutions in the construction and operation of the instrument.

%
%
%

\vspace{0.5cm}

\noindent \textbf{Note:} Collaborations have the possibility to provide an authors list in xml format which will be used while generating the DOI entries making the full authors list searchable in databases like Inspire HEP. For instructions please go to icrc2021.desy.de/proceedings or contact us under icrc2021proc@desy.de.

\end{document}